\documentclass[12pt]{article}
\title{Strong coupling constant from bottomonium \\
fine structure.}
\vspace{2cm}
\author{A.M.Badalian$^a$ and B.L.G.Bakker$^b$\\
\\
$^a$ Institute of Theoretical and Experimental Physics,\\
117218, Moscow, B.Cheremushkinskaya, 25, Russia \\
$^b$ Department of Physics and Astronomy,\\
Vrije Universiteit, Amsterdam, The Netherlands}
\date{}
\begin{document}

\maketitle

\def\la{\mathrel{\mathpalette\fun <}}
\def\ga{\mathrel{\mathpalette\fun >}}
\def\fun#1#2{\lower3.6pt\vbox{\baselineskip0pt\lineskip.9pt
\ialign{$\mathsurround=0pt#1\hfil##\hfil$\crcr#2\crcr\sim\crcr}}}

\vspace{1cm}

\begin{abstract}
From a fit to the experimental data on the $b \bar{b}$ fine structure,
the two-loop strong coupling constant is extracted. For the $1P$ state
the fitted value is $\alpha_s (\mu_1) = 0.33 \pm 0.01\,({\rm exp}) \pm
0.02\, ({\rm th})$ at the scale $\mu_1 = 1.8 \pm 0.1\; {\rm GeV}$, which
corresponds to the QCD constant $\Lambda^{(4)}\, ({\rm 2-loop}) = 338 \pm
30\; {\rm MeV} ~ (n_f = 4)$ and $\alpha_s(M_Z) = 0.119 \pm 0.002$. For
the $2P$ state the value $\alpha_s(\mu_2) = 0.40 \pm 0.02\, ({\rm exp})
\pm 0.02\,({\rm th})$ at the scale $\mu_2 = 1.02 \pm 0.02 \; {\rm GeV}$
is extracted, which is essentially larger than in the previous analysis
of refs. \cite{4,5}, but about $30 \%$ smaller than the value given by the
standard perturbation theory. This value $\alpha_s(1.0) \approx 0.40$ can
be obtained in the framework of the background perturbation theory thus
demonstrating the freezing of $\alpha_s(\mu)$. The relativistic
corrections to $\alpha_s$ are found to be about $15 \%$.
\end{abstract}


\newpage

\section{Introduction \label{sec.1}}

The bottomonium spectrum is one of the richest among all known mesons and
its levels were measured with high precision \cite{1}. These
data about $b \bar{b}$ states have been intensively studied in
different theoretical approaches, in particular, to determine the QCD
strong coupling constant $\alpha_s(\mu)$ at different energy scales
$\mu$ from the level differences [2-10]. At present, however, there is
no clear picture which are the exact values of $\alpha_s(\mu)$ for the $b
\bar{b}$ levels and how they are changing from the ground state to the
excited ones. There are several reasons for that.

First of all, there is no experimental information on the $\eta_b(n S)$
masses and therefore $\alpha_s(\mu)$ cannot be
directly determined from the $b \bar{b}$ hyperfine splittings
in S-wave states.

Second, to describe the fine structure splittings in the  $P$-wave
states, different energy scales $\mu$ were used in different
theoretical analyses [4-7]. In ref.~\cite{4} $\alpha_s(\mu) = 0.33 \;
(\mu = 3.25\; {\rm GeV})$ was taken for all $b \bar{b}$ $S$- and
$P$-wave states, while in \cite{5} $\mu$ was chosen to be equal to the
$b$ quark mass, $\mu = m$ with either $m = 4.6$ GeV or $m = 5.2$ GeV.  
The fitted values of $\alpha_s(\mu)$ were found to be
$\alpha_s(m) = 0.22 \div 0.27$ \cite{5} and for the $2P$ state
$\alpha_s(\mu)$ appeared to be smaller than for the $1P$ state.

An important step to clarify this problem was taken in \cite{6,7} where the
low-lying bottomonium states, $1S$, $2S$, and $1 P$ were investigated. It was
observed there that the scale $\mu$ is a decreasing function of the
principal quantum number $n$,$\mu = 2 (na)^{-1}$ where $a$ is a
Coulomb type radius. Therefore, $\mu$ is found to be equal for the $2S$ and $1P$
states and the values $\mu = 1.7 GeV$, $\alpha_s(1.7)=0.29$, were determined from the fine
structure splittings of $\chi_b(1P)$. Also, $\alpha_s(\mu)$ is larger
for excited states with a larger radius of the system, thus indicating
that for a bound state the characteristic scale $\mu$ is determined by
the size, but not by the momentum of the system. One of our main goals here
is to check this important statement for the $2P$ state, $\chi_b(2P)$,
which cannot be studied in the framework of the approach developed in
\cite{6,7}.

In the present study of the $1P$ and $2P$ $b \bar{b}$ states we shall try to answer the
following questions:

\begin{quote}
What are the values of $\alpha(\mu)$ for the $2P$ and the $1P$ states?

Do the extracted values of $\alpha_s(\mu)$ correspond to
the existing experimental data on $\alpha_s(M_Z)$ and $\Lambda^{(n_f)}$?

How does $\alpha_s(\mu)$ depend on the relativistic corrections to the wave
functions in bottomonium?

How sensitive are the extracted values of $\alpha_s(\mu)$ to the $b$
quark pole mass and the parameters of the static interaction?
\end{quote}

\section{Perturbative Radiative Corrections \label{sec.2}}

It is well known that one cannot describe the fine structure splittings
in heavy quarkonia without taking into account the second order
radiative corrections [4-7, 10].  In coordinate space, perturbative
static and spin-dependent potentials in the $\overline{MS}$ 
renormalization scheme were obtained in \cite{2,3}. From the potentials given
there one can  immediately find the matrix elements of the spin-orbit and the
tensor potentials: $a = \langle {V}_{LS} (r) \rangle$, $c = \langle
{V}_T (r) \rangle$. Below we give their expressions for a number of
flavours $n_f = 4$, valid for the $b \bar{b}$ system:
\begin{equation}
a_{\rm P} = a^{(1)}_{\rm P} + a^{(2)}_{\rm P},
 \label{eq.01}
\end{equation}
\begin{equation}
 a^{(1)}_{\rm P} = \frac{2 \alpha_s(\mu)}{m^2} \langle r^{-3} \rangle,  \quad 
 a^{(2)}_{\rm P} =
 \frac{2 \alpha^2_s(\mu)}{\pi m^2} \left\{ \langle r^{-3} \rangle
 \left( \frac{25}{6} \ln \left(\frac{\mu}{m}\right) + A \right) +
\frac{13}{6} \langle r^{-3} \ln mr \rangle \right\}
 \label{eq.02}
\end{equation}
and for the
perturbative part of the the tensor splitting $c_{\rm P}$,
\begin{equation}
c_{\rm P} = c^{(1)}_{\rm P} + c^{(2)}_{\rm P},
 \label{eq.03}
\end{equation}
\begin{equation}
c^{(1)}_{\rm P} = \frac{4}{3}\;\frac{\alpha_s(\mu)}{m^2} \langle r^{-3} \rangle, \quad
c^{(2)}_{\rm P} = \frac{4}{3}~ \frac{\alpha^2_s (\mu)}{\pi m^2} \left
\{\langle r^{-3} \rangle \Biggl (\frac{25}{6} \ln \frac{\mu}{m} + B\Biggr ) +
\frac{7}{6} \langle r^{-3} \ln mr \rangle \right \}.
 \label{eq.04}
\end{equation}
Here the constant $A = \frac{13}{6} \gamma_E + \frac{7}{36} = 1.44508$ and
$B = \frac{7}{6} \gamma_E + \frac{33}{12} = 3.42342$.

For our analysis it is convenient to introduce a linear
combination of the matrix elements $a$ and $c$ as was done in ref.~\cite{10}:
$\eta = \frac{3}{2} c - a$. Its perturbative part $\eta_{\rm P}$ is
\begin{equation}
\eta_{\rm P} = \frac{3}{2} c_{\rm P} - a_{\rm P} = \frac{3}{2} c^{(2)}_{\rm P} - a^{(2)}_{\rm P} =
\frac{2 \alpha^2_s(\mu)}{\pi m} f_4 .
 \label{eq.05}
\end{equation}
The factor $f_4$ in Eq.~(5)  can be found from Eqs.~(2) and (4),
\begin{equation}
f_4(n P) = \frac{1}{m} \left[ 1.97834\; \langle r^{-3} \rangle_{nP} - \langle
r^{-3} \ln mr \rangle_{nP}\right] .
 \label{eq.06}
\end{equation}
For the fine structure analysis it turns out to be very important that
the combination of matrix elements $f_4$ does not depend on the energy
scale $\mu$. Later, it will be also shown that $f_4$ has the largest
relativistic correction (about $35\%$) compared to other matrix
elements and depends weakly on the parameters of the static interaction and
on the mass of the $b$-quark.

\section{Nonperturbative Contributions \label{sec.3}}

Besides the perturbative terms Eqs.~(\ref{eq.02}, \ref{eq.04}) the tensor
and spin-orbit splittings have in general nonperturbative
contributions: $a = a_{\rm P} + a_{\rm NP}$ and $c = c_{\rm P} + c_{\rm NP}$.

The nonperturbative part of the spin-orbit parameter $a_{\rm NP}$ is
determined by the chosen confining potential. Here we shall take a
linear potential $\sigma r$ at all distances. The linear behaviour of
the confining potential at large distances is well established
phenomenologically due to the existence of the meson Regge trajectories
and was also deduced from the minimal area law for Wilson loops in QCD
(see the review \cite{11} and the references therein). Recently, the
linear behaviour of the confining potential was found at small
distances due to the interferences of perturbative and nonperturbative
(NP) effects \cite{12} and to the saturation property of the QCD strong
coupling constant in vacuum field [13-15]. For the linear potential
$\sigma r$ the nonperturbative interaction is given by the Thomas
potential for which
\begin{equation}
 a_{\rm NP} = - \frac{\sigma}{2m^2} \langle r^{-1} \rangle
 \label{eq.07}
\end{equation}
The nonperturbative contribution to the tensor splitting can be found from the 
vacuum field correlator $D_1(x)$ \cite{10, 16} which was measured in lattice 
QCD \cite{17} and was found to be of exponential form. Then, as was shown in
\cite{10, 18, 19},
\begin{equation}
c_{\rm NP} = \frac{D_1(0)}{3 m^2 T_g} \langle r^2 K_0 (r/T_g) \rangle
 \equiv \frac{D_1 (0)}{3 m^2} J~ (T_g), \quad J(T_g) \equiv
\frac{1}{T_g} \langle r^2 K_0 (r/T_g) \rangle,
 \label{eq.08}
\end{equation}
where $T_g$ is the vacuum correlation length. Lattice QCD without
dynamical fermions give $T_g \approx 0.2$ fm and $T_g \approx 0.3$ fm in the
presence of dynamical fermions with four flavours\cite{17}.

In refs.~\cite{17} the correlator $D_1(0)$
in Eq.~(\ref{eq.08}) was shown to be small: lattice calculations in quenched SU(3)
theory give $D_1(0)/D(0) \approx \frac{1}{3}$ and in full QCD
with four staggered fermions $D_1(0)/D(0) \approx 0.1$, where  $D(0)$
is another vacuum field correlator which mostly determines the confining
potential. These two correlators  at the point $x = 0$ can be expressed
through the vacuum gluonic condensate $G_2$ (here the vacuum correlators
are normalized as in \cite{11,16}):
\begin{equation}
D(0)+ D_1(0) = \frac{\pi^2}{18} ~ G_2.
 \label{eq.09}
\end{equation}
Therefore, the lattice estimate for $D_1(0)/D(0)$ is $0.1 \div 0.3$ and from
the relation (9) one obtains
\begin{equation}
\frac{\pi^2}{180} G_2 \la D_1 (0) \la \frac{\pi^2}{72}~ G_2 .
 \label{eq.10}
\end{equation}
Our calculations give the following typical values for the
matrix
elements $J(T_g)$: for the $1P$ state $J(T_g) \approx 0.17\; {\rm GeV}^{-1}$
and for the $2P$ state $J(T_g) \approx 0.20\; {\rm GeV}^{-1}$, if the $b$ quark
mass $m \approx 4.8$ GeV and $T_g \approx 0.2 \div 0.3$ fm is taken.
Then, if the value of the gluonic condensate $G_2 = 0.05 \pm 0.02 \;
{\rm GeV}^4$ \cite{7} is used, one finds the estimate in quenched QCD
\begin{equation}
c_{\rm NP} \la 0.03 \pm 0.01 \;{\rm MeV}.
 \label{eq.11}
\end{equation}
In full QCD an even smaller value is found. This value of $c_{\rm NP}$ is
much less than both $\vert a_{\rm NP} \vert$, Eq.~(7), and the experimental
errors. Therefore it can be neglected in the tensor splitting
$c$ and also in $\eta_{\rm NP} = \frac{3}{2} c_{\rm NP} - a_{\rm NP}$,
i.e.we take here $\eta_{\rm NP} = \-a_{\rm NP}$.

\section{Fitting Conditions \label{sec.4}}

To fit the experimental data
\begin{eqnarray}
 a_{\rm exp} (1P) & = & 14.23 \pm 0.53 \;{\rm MeV}, \;
 a_{\rm exp} (2P)  =  9.39 \pm 0.18 \;{\rm MeV},
 \nonumber \\
 c_{\rm exp} (1P) & = & 11.92 \pm 0.25 \;{\rm MeV}, \;
 c_{\rm exp}(2P) = 9.14 \pm 0.25\; {\rm MeV}, 
 \label{eq.12}
\end{eqnarray}
the following conditions have to be satisfied:
\begin{eqnarray}
a_{\rm tot}(n P) & = & a^{(1)}_{\rm P} + a^{(2)}_{\rm P} - \frac{\sigma}{2m^2} \langle r^{-1} \rangle =
a_{\rm exp} (nP)
 \nonumber \\
c_{\rm tot}(nP) & = & c^{(1)}_{\rm P} + c^{(2)}_{\rm P} = c_{\rm exp} (nP) .
 \label{eq.13}
\end{eqnarray}

As seen from Eqs.~(2) and (4), the l.h.s. of these expressions strongly
depend on the normalization scale $\mu$, but the combination $\eta$ does not. 
The fitting condition for $\eta$ is
\begin{equation}
\eta(nP) = \frac{2 \alpha^2_s (\mu)}{\pi m} f_4 + \frac{\sigma}{2 m^2}
\langle r^{-1} \rangle = \eta_{\rm exp},
 \label{eq.14}
\end{equation}
where the experimental values for $\eta^{(nP)}_{\rm exp}$ are
\begin{equation}
\eta_{\rm exp}(1P)  =  3.65 \pm 0.9\; {\rm MeV}, \quad
\eta_{\rm exp} (2P) =  4.32 \pm 0.4\; {\rm MeV} .
 \label{eq.15}
\end{equation}
The condition (\ref{eq.14}) does not depend explicitly on $\mu$ and can be
rewritten as
\begin{equation}
\frac{2 \alpha^2_s (\mu)}{\pi m} f_4 (nP) = \eta_{\rm exp} -
\frac{\sigma}{2m^2} \langle r^{-1} \rangle \equiv \Delta (nP),
 \label{eq.16}
\end{equation}
hence  the strong coupling constant can be expressed as
\begin{equation}
\alpha_s(\mu) = \sqrt{\frac{\pi m \Delta}{2 f_4}} .
 \label{eq.17}
\end{equation}
For a chosen interaction and quark mass $m$, $\Delta(nP)$ and $f_4(nP)$ are
known numbers and one can immediately determine $\alpha_s(\mu)$.

The extraction of $\alpha_s(\mu)$ from the condition (\ref{eq.17}), in general,
extremely simplifies the fit and also puts strong restrictions on the
possible choice of the normalization scale $\mu$. Just this condition
was exploited in \cite{10} to determine $\alpha_s(\mu)$ for the $1P$ state in
charmonium. In charmonium $\eta _{exp}\approx 24\; {\rm MeV}$ and the
typical value
of $\Delta \approx 7 \div 8$ MeV is not small, so the uncertainty
in the extracted value of $\alpha_s(\mu)$ is about $10\%$.

In bottomonium the typical values of $\vert a_{\rm NP} \vert$ are found to
be smaller: 
$\vert a_{\rm NP} (1P) \vert = 2.6 \pm 0.2$ MeV (see Table~\ref{tab.05})
and $\vert a_{\rm NP} (2P) \vert = 1.95 \pm 0.10$ MeV (see Table~\ref{tab.04}).
As a result, the numerical values of $\Delta(nP)$ to be substituted in
Eq.~(\ref{eq.17}) are small:
\begin{equation}
\Delta(1P)  =  1.05 \pm 0.9 ({\rm exp}) \pm 0.15 ({\rm th}) \; {\rm MeV}, \quad
\Delta(2P)  =  2.4 \pm 0.4 ({\rm exp}) \pm 0.10 ({\rm th})\; {\rm MeV} .
 \label{eq.18}
\end{equation}
The theoretical uncertainties in this equation are caused by the uncertainty
of the value of $a_{\rm NP}$ in the Thomas interaction.
Still, for the $2P$ state the total error in $\Delta (2P)$ is not large,
about $20\%$, and therefore $\alpha_s(\mu)$, proportional to
$\sqrt{\Delta}$, can be determined from the condition (\ref{eq.17}) with an accuracy of
about $10\%$. Our calculations show also that the matrix element $f_4$
in Eq.~(\ref{eq.17}) is practically constant and therefore the theoretical error in
Eq.~(\ref{eq.18}) coming from $f_4$ is small.

For the $1P$ state the experimental error in $\eta_{\rm exp}$,
 Eq.~(\ref{eq.15}), as well as in
$\Delta$, Eq.~(\ref{eq.18}), is large: it comes mostly from the experimental
uncertainty in the $\chi_{b_0}(1P)$ mass. Therefore $\Delta(1P)$ can vary
in a wide range: $0 \leq \Delta \leq 2.0$ MeV and the relation (\ref{eq.17})
cannot give an accurate value for $\alpha_s(\mu)$. Instead, for the $1P$ state
one needs to use the conditions (\ref{eq.13}) which are $\mu$-dependent and less
restrictive.

\section{Dependence on Scale \label{sec.5}}

The second-order perturbative corrections to the spin-orbit and tensor
splittings, which are not small, explicitly depend on the scale $\mu$.
In Eqs.~(\ref{eq.02}, \ref{eq.04}) $\ln (\mu/m)$ enters with the large
coefficient 25/6 and therefore the choice $\mu = m$ (causing this
logarithm to vanish) can give rise to inconsistent results. Just this
choice was taken in \cite{5} where two $b$-quark masses, $m = 4.6$ GeV and
$m = 5.2$ GeV, were analysed. We shall discuss here some results of
ref.~\cite{5}.

From the fit in \cite{5} it was obtained that the value
$\tilde{\alpha_s} (m)$ extracted from the tensor and the spin-orbit
splittings are
slightly different and for the $2P$ state this difference is
increasing.  (Here $\tilde{\alpha} (\mu)$ or $\tilde{\alpha}
(\tilde{\mu})$ denotes the fitted (extracted) value of the strong
coupling  constant.)

Also, for the $2P$ state $\tilde{\alpha} (5.2) = 0.26 \pm 0.01$ is a bit
larger than $\tilde{\alpha}_s (4.6) = 0.25 \pm 0.01$ for the smaller
$b$-quark mass, in contradiction with the standard behaviour of the
running coupling constant $\alpha_s(q^2)$.

The extracted value, $\tilde{\alpha}_s (m) \approx 0.25 \div 0.27$,
turned out to be about $20\%$ larger than the values $\alpha_s(4.6)$
and $\alpha_s(5.2)$  calculated with the conventional value of
$\Lambda^{(4)}$ Eq.~(\ref{eq.19}):  $\alpha_s(4.6) = 0.22 \pm 0.01$, 
$\alpha_s(5.2) = 0.21 \pm 0.01$.

In the calculations that follow, it will be easy to compare our
results with those from ref.~\cite{5} because in both cases the same
perturbative interaction and linear potential $\sigma r$ were used.
However, the calculations of ref.~\cite{5} were done in the
nonrelativistic case (for fixed $\sigma = 0.2\; {\rm GeV}^2$ and two
$b$-quark masses). Here both relativistic and nonrelativistic cases
will be considered and $\sigma$, $m$, and $\alpha_{\rm eff}$ of the
Coulomb potential will be varied in a wide range.  From our analysis it
will be clear that the inconsistencies in the $\tilde{\alpha}_s (\mu)$
behaviour mentioned above, are related to the a priori choice $\mu = m$
made in \cite{5}.

At this point it is worthwhile to note that at present the QCD constant
$\Lambda^{(n_f)}$ is well known for $n_f = 5$, because $\alpha_s(M_z) =
0.119 \pm 0.002$ is established from different experiments:
$\Lambda^{(5)} (2-{\rm loop}) = 237^{+26}_{-24}$ MeV and $\Lambda^{(5)}
(3-{\rm loop}) = 219^{+25}_{-23}$ MeV are given in \cite{1}. Then from
the matching of $\alpha_s(\mu)$ at the scale $\mu = \bar{m}_b$
($\bar{m}_b$ is the running mass in the $\overline{MS}$ scheme)  and
taking $\bar{m}_b = 4.3 \pm 0.2$ GeV \cite{1} one can find
$\Lambda^{(4)} (3-{\rm loop}) = 296^{+31}_{-29}$ MeV or in the two-loop
approximation $\Lambda^{(4)}$ is
\begin{equation}
\Lambda^{(4)} (2-{\rm loop}) = 338^{+33}_{-31}\; {\rm MeV}.
 \label{eq.19}
\end{equation}
It is of interest to compare $\alpha_s(\mu)$ for $\Lambda^{(4)}$ given
by Eq.~(\ref{eq.19}) with the fitted values $\tilde{\alpha}_s
(\tilde{\mu})$ used in different theoretical analyses: $a_s (3.25) =
0.251 \pm 0.009$ whereas in \cite{4} the fitted value $\tilde{\alpha}_s
(3.25) = 0.33; ~~ \alpha_s (4.60) = 0.221 \pm 0.007$ while in \cite{5}
$\tilde{\alpha}_s (4.6) \approx 0.27$. In both fits the extracted
values appeared to be about $20\%$ larger.

This $20\%$ difference implies either very large values of
$\Lambda^{(4)}$ or an essentially smaller scale of $\mu$. For example,
$\alpha_s(\mu_0) = 0.33$ with the conventional
$\Lambda^{(4)}$, Eq.~(\ref{eq.19}), corresponds to $\mu_0 = 1.80 \pm
^{0.18}_{0.16}$ GeV instead of $\tilde{\mu} = 3.25$ GeV in \cite{4} and
this $\mu_0$ would be in  good agreement with the one cited in
\cite{6,7} and with our result (see Section \ref{sec.9}).

In our present analysis when different sets of parameters are taken, we
shall impose two additional restrictions:

1) For the given $P$-state the extracted value of $\tilde{\alpha}_s (\mu)$ must
be the same for the tensor and the spin-orbit splittings, because both
interactions have the same $r^{-3}$ behaviour and they also
have the same characteristic size (momentum).

2) Only those sets of parameters for which the fitted two-loop value of
$\tilde{\alpha}(\mu)$ corresponds to the conventional value of 
$\Lambda^{(4)}$ in two-loop approximation,  Eq.~(\ref{eq.19}), are deemed  appropriate.

\section{Static Potential \label{sec.6}}

In heavy $Q \bar{Q}$ systems the spin-dependent interaction contains
the factor $m^{-2}$ and therefore it is small and can be considered as
a perturbation.  For the unperturbed Hamiltonian we considered two
cases,relativistic and nonrelativistic,
\begin{equation}
H^R_0 = 2~ \sqrt{\vec{p}^{\, 2} + m^2} + V_{\rm st} (r)
 \label{eq.20}
\end{equation}
or
\begin{equation}
H^{NR}_0 = \frac{\vec{p}^{\,2}}{m} + V_{\rm st} (r) .
 \label{eq.21}
\end{equation}

Here a static potential, $V_{\rm st} (r) = V^P_{\rm st} (r) + V^{NP}_{\rm st} (r)$,
needs some remarks. The perturbative static potential is now known in
two-loop approximation \cite{20}, but for our discussion it is enough to take
it in one-loop approximation from \cite{3}:
\begin{equation}
V^P_{\rm st} = -\frac{4}{3} \frac{\alpha_V(r)}{r}
 \label{eq.22}
\end{equation}
Here the vector coupling constant $\alpha_V(r)$ is expressed through
$\alpha_s(\mu)$ in the $\overline{MS}$ scheme in the following way \cite{3}:
($\alpha_s(\mu)<<1$)
\begin{eqnarray}
 \alpha_V(r) & = & \alpha_s(\mu) \Biggl [1 + \frac{\alpha_s(\mu)}{\pi}
 \left(a_1+\frac{\beta_0}{2}~(\ln(\mu r)+\gamma_E) \right) \Biggr ]
 \nonumber \\
 & = & \frac{\alpha_s(\mu)}
 {1-\frac{\alpha_s(\mu)}{\pi}\left(a_1+\frac{\beta_0}{2}
 (\ln(\mu r)+\gamma_E)\right)}
 \to \frac{4\pi}{\beta_0 \ln ((\Lambda_R r)^{-2})}.
 \label{eq.23}
\end{eqnarray}
In Eq.~(\ref{eq.23}) we have used 
$\alpha_s(\mu)= 4\pi/[\beta_0 \ln(\mu^2 / \Lambda^2_{\overline{MS}})]$,
and the conventional QCD constant in coordinate space: 
$\Lambda_R=\Lambda_{\overline{MS}} \exp(\gamma_E+a)$ where
$a=2 a_1/\beta_0$.  We see that the dependence on $\mu$ disappears.
The constants are: $\beta_0=11-2 n_f/3$, so for $n_f=4$, $\beta_0=25/3$;
$a_1=31/12-5 n_f/18$, so for $n_f=4$, $a_1=53/36$.

This expression is valid only for small radiative corrections or
small distances: $r e^{\gamma_E} \tilde{\Lambda}^{(4) } \ll 1$ or $r \ll 2
\;{\rm GeV}^{-1} = 0.4$ fm $(\tilde{\Lambda}^{(4)} \approx 0.3$ GeV).
However, in
bottomonium the sizes of the different states are varying in a wide range,
e.g., typical values of the root-mean-square radius, $R(n L) = \sqrt{\langle
r^2 \rangle _{nL}}$, are:
\begin{eqnarray}
R(1S) & = & 0.2 \; {\rm fm}, \; R(1P) = 0.4 \; {\rm fm}, \; R(2S) = 0.5 \; {\rm fm}, R(2P) = 0.65 \; {\rm fm},
 \nonumber \\
R(3S) & = & 0.7 \; {\rm fm}, \; R(3P) = 0.85 \; {\rm fm}, \; R(4S) = 0.9 \; {\rm fm}.
 \label{eq.24}
\end{eqnarray}
These numbers are practically independent of the choice of the static
potential parameters and the confining potential provided the chosen
potential reproduces the bottomonium spectrum with good accuracy.

From Eq.~(\ref{eq.24}) one can see that the sizes of the $nL$ states
run from 0.2 fm to 0.9 fm. Therefore the perturbative potential,
Eq.~(\ref{eq.23}), valid for $r \ll 0.4$ fm, can be used only for
low-lying states. For the $1S$, $2S$, and $1P$ states this perturbative
interaction (also with two-loop corrections) was analyzed in detail in
refs.~\cite{6,7} and there it was found that (i) for the $1S$ and $2S$
states the values of $\mu$ are different and (ii) $\mu$ is smaller in
the  $2S$ state.  Therefore, one can expect that for every level a
specific consideration is needed to determine $\mu$ or $\alpha_s(\mu)$.

To describe the $2P$ state, the size of which is about 0.65 fm, or the $b
\bar{b}$ spectrum as a whole, a different approach is needed. Here
we suggest instead of the perturbative potential Eq.~(\ref{eq.22}) to
use the perturbative potential in  background vacuum field,
$V_B(r)$:
\begin{equation}
V_B(r) = -\frac{4}{3}~ \frac{\alpha_B (r)}{r},
 \label{eq.25}
\end{equation}
in momentum space
\begin{equation}
V_B(q^2) = -\frac{4}{3}~ \frac{4 \pi}{\vec{q}^{\,2}} ~ \tilde{\alpha}_B
(q^2), \quad q^2 \equiv \vec{q}^{\, 2}.
 \label{eq.26}
\end{equation}
In this potential $\tilde{\alpha}_B (q^2)$ is a vector coupling
constant in vacuum background field which was introduced in \cite{14} and
applied to $e^+e^- \to$ hadrons processes in \cite{21}:
\begin{equation}
 \tilde{\alpha}_B(q^2) = \frac{4 \pi}{\beta_0 t_B}
 \left[1-\frac{\beta_1 \ln t_B(q)}{\beta^2_0 t_B(q)} \right] , \quad
 t_B(q) = \ln\, \frac{q^2 + m^2_B}{\tilde{\Lambda}^2},
 \label{eq.27}
\end{equation}
with $\beta_0 = 25/3$.
For the vector coupling constant, $\alpha_V(q^2)$, $\tilde{\Lambda}$
differs from $\Lambda$ in the $\overline{MS}$ scheme:  $\tilde{\Lambda}
= \Lambda^{(4)}_{\overline{MS}} ~ e^a = 481^{+47}_{-41}$ MeV, $a =
\frac{5}{6} - \frac{4}{\beta_0} = 0.35333$, and $\Lambda^{(4)}_{\overline{MS}}$
was
taken from Eq.~(\ref{eq.19}). (In the $\overline{MS}$ scheme
$\Lambda_B$ and $\Lambda_{\overline{MS}}$ coincide for $n_f = 4,5$
because of their identical behaviour at large $q^2$ \cite{10}.) The
background mass $m_B$ was found from the fit to the charmonium
fine structure in \cite{10} where $m_B = 1.1$ GeV was obtained.

In coordinate space $\alpha_B(r)$ can be calculated from the Fourier
transform of the potential Eq.~(\ref{eq.26}) with $\alpha_B(q^2)$ given by 
Eq.~(\ref{eq.27}). Then
\begin{equation}
 \alpha_B(r) = \frac{8}{\beta_0}~ \int \limits^{\infty}_{0} dq~
 \frac{\sin qr}{q t_B(q)}~ \left[ 1 - \frac{\beta_1}{\beta^2_0}~
 \frac{\ln t_B(q)}{t_B(q)} \right].
 \label{eq.28}
\end{equation}
The strong
coupling constant in vacuum background field maintains the property of
asymptotic freedom at small $r$, $r \ll \tilde{\Lambda}^{-1}$ and $r \ll
m^{-1}_B$,
\begin{equation}
\alpha_B(r \to 0) = -\frac{2 \pi}{\beta_0 \ln(\tilde{\Lambda} e^{\gamma} r)} \,
 \label{eq.29}
\end{equation}
Here  the function $\gamma=\gamma(r)$ is
\begin{equation}
\gamma = \gamma(r) = \gamma_E + \Sigma, \quad
 \Sigma = \sum \limits^{\infty}_{k=1} ~ \frac{(-m_B r)^k}{k! k},
 \label{eq.30}
\end{equation}
or at small $r$
\begin{equation}
\gamma = \gamma_E - m_B r ,
 \label{eq.31}
\end{equation}
whereas in standard perturbative theory $\gamma^P = \gamma_E =
0.5772$. Due to the dependence  on the distance $r$ in
Eq.~(\ref{eq.31}) the expression Eq.~(\ref{eq.29}) is always bounded.

For large $r^2$, $r^2 \gg m^{-2}_B$, the limit of $\alpha_B(r)$ in
Eq.~(\ref{eq.28}) tends to a constant, denoted as $\alpha_B(\infty)$ and
called the freezing value:
\begin{equation}
\alpha_B(\infty) = \frac{4 \pi}{\beta_0 t_0}
 \left[ 1 - \frac{\beta_1 \ln t_0}{\beta^2_0 t_0} \right], \quad
 \quad t_0 = \ln \, \frac{m^2_B}{\tilde{\Lambda}^2}.
 \label{eq.32}
\end{equation}
From the integral Eq.~(\ref{eq.28}) it can be shown that the freezing
value is the same in coordinate and  in momentum space,
$\alpha_B(r\to\infty) = \tilde{\alpha}_B(q^2=0)$.  The properties
of $\alpha_B(r)$ were discussed in \cite{10,13,14} and a detailed analysis
of $\alpha_B(r)$ will be published elsewhere.

In the intermediate region, $0.2 \; {\rm fm} \leq r \leq 0.9$ fm, $\alpha_B(r)$
approaches rapidly the value $\alpha_B(\infty)$.

Therefore, to study the bottomonium spectrum as a whole it is convenient to
introduce an effective constant $\alpha_{\rm eff}$:
\begin{equation}
 \alpha_B(r) = \alpha_{\rm eff} + \delta \alpha_B(r),\quad
 \alpha_{\rm eff} = {\rm constant}, \quad
 \quad\vert \delta \alpha_B(r) \vert \ll  \alpha_{\rm eff},
 \label{eq.33}
\end{equation}
and to consider the contribution from the term $\delta V_B(r)$,
\begin{equation}
\delta V_B(r) = -\frac{4}{3}~ \frac{\delta \alpha_B(r)}{r},
 \label{eq.34}
\end{equation}
as a perturbation.
Then in the Hamiltonian (22) the static interaction
\begin{equation}
V_0(r) = -\frac{4}{3}~ \frac{\alpha_{\rm eff}}{r}
 \label{eq.35}
\end{equation}
will be taken into account as an unperturbed interaction.

For the nonperturbative interaction a linear form $\sigma r$ will be taken and
therefore the static potential in the unperturbed Hamiltonian $V_0(r)$,
\begin{equation}
V_0(r) = -\frac{4}{3}~ \frac{\alpha_{\rm eff}}{r} + \sigma r + C_0
 \label{eq.36}
\end{equation}
coincides with the well known Cornell potential. Later, the values
of the string tension $\sigma$ will be varied in the range $0.17 \div 0.20\; {\rm GeV}^2$.

We shall present a detailed analysis of the $b \bar{b}$ spectrum in a
separate paper.

\section{Relativistic Corrections \label{sec.7}}

There  exists the point of view that in bottomonium the relativistic
corrections are small because of the heavy $b$ quark mass. Indeed, the
comparison of levels and mass differences for the Schr\"odinger
equation and the Salpeter equation, Eqs.~(\ref{eq.20},\ref{eq.21}), in general,
confirms this statement (here the  static potential is supposed to be
the same in both cases). In Table \ref{tab.01} the $b\bar{b}$ mass
differences are given for two typical sets of parameters.

\begin{table}
\caption[]{Bottomonium level differences (MeV) for the Schr\"odinger and the
 Salpeter equations.}
\label{tab.01}
\vspace{1ex}
\begin{center}
\begin{tabular}{|l|r|r|r|r|r|}
\hline
 Mass differences 
 & \multicolumn{2}{c|}{Set I, $\alpha_{\rm eff} = 0.3545$} &
   \multicolumn{2}{c|}{Set II, $\alpha_{\rm eff} = 0.36$} & Exp. val. \\
 & \multicolumn{2}{c|}{$m=4.737$ GeV } &
   \multicolumn{2}{c|}{$m=4.81$ GeV } &  (MeV) \rule{1mm}{0mm}\\
 & \multicolumn{2}{c|}{$\sigma =0.20$ GeV$^2 \; ({}^a)$} &
   \multicolumn{2}{c|}{$\sigma =0.18$ GeV$^2 $} &  \\
\hline
 & Rel.\rule{5mm}{0mm}  & Nonrel.    & Rel.\rule{5mm}{0mm} & Nonrel.    &  \\
\hline
 $M(2S) - M(1S)$ & 554.34 & 551.97 & 556.55 & 550.03 & $562.9\pm0.5$ \\
 $M(3S) - M(2S)$ & 350.43 & 354.78 & 335.62 & 338.49 & $332.0\pm0.8$ \\
 $M(4S) - M(3S)\; ({}^b)$ & 285.93 & 291.83 & 270.63 & 275.30 & $224.7\pm4.0$ \\
 $M(1P) - M(1S)$ & 458.04 & 439.66 & 473.49 & 450.15 & $439.8\pm0.9$ \\
 $M(2P) - M(1P)$ & 359.67 & 366.75 & 342.55 & 348.70 & $359.8\pm1.2$ \\
 $M(2S) - M(1P)$  & 96.31   & 112.31  & 83.07  & 99.88   & $123.1\pm1.0$\\
 $M(3S) - M(2P)$ &  87.06 & 100.34 & 72.82  & 89.67   & $95.3\pm1.0$\\
\hline
\end{tabular}
\end{center}

 $({}^a)$ This set was taken from ref.~\cite{22}\\
 $({}^b)$ The $4S$ level lies above the $B\bar{B}$ threshold
\end{table}

From Table \ref{tab.01} one can see that

\begin{quote}
(i) Relativistic corrections are small for large mass differences like
$M(n,L)-M(n-1,L)$ or $M(nL)-M(n,L-1)$;

(ii) For close lying levels, like $\Delta_1=M(2S)-M(1P)$  and
$\Delta_2=M(3S)-M(2P)$, the corrections are essential, about 15\%, and to get
$\Delta_1$ and $\Delta_2$  close to the experimental data it is
necessary to take into account the contribution from the perturbation
$\delta V_B(r)$ Eq.~(\ref{eq.34}). In the relativistic case the
influence of the phenomenon of asymptotic freedom appears to be more
essental than in the nonrelativistic (NR) case.
\end{quote}

\begin{table}
\caption[]{$1P$-state matrix elements for the Schr\"odinger and the
 Salpeter equations.}
\label{tab.02}
\vspace{1ex}
\begin{center}
\begin{tabular}{|l|r|r|r|r|}
\hline
 Matrix element 
 & \multicolumn{2}{c|}{Set I $({}^a)$} &
   \multicolumn{2}{c|}{Set II $({}^a)$} \\
\hline
 & Rel.\rule{5mm}{0mm}  & Nonrel.    & Rel.\rule{5mm}{0mm} & Nonrel.  \\
\hline
 $\surd\langle r^2 \rangle$ (GeV${}^{-1}$)  & 1.994 & 2.039 & 2.008 & 2.054  \\
 $\langle r^{-1} \rangle$ (GeV)  & 0.633 & 0.614 & 0.631 & 0.612  \\
 $\langle r^{-3} \ln mr \rangle$ (GeV${}^3$) & 0.675 & 0.631 & 0.681 & 0.636  \\
 $\langle r^{-3} \rangle$ (GeV${}^3$)  & 0.551 & 0.483 & 0.556 & 0.485  \\
 $f_4(1P)$ (GeV${}^2$) & 0.0876 & 0.0685 & 0.0871 & 0.0673  \\
\hline
\end{tabular}
\end{center}

 $({}^a)$ For the parameters see Tab.~\ref{tab.01}
\end{table}

\begin{table}
\caption[]{$2P$-state matrix elements for the Schr\"odinger and the
 Salpeter equations.}
\label{tab.03}
\vspace{1ex}
\begin{center}
\begin{tabular}{|l|r|r|r|r|}
\hline
 Matrix element 
 & \multicolumn{2}{c|}{Set I $({}^a)$} &
   \multicolumn{2}{c|}{Set II $({}^a)$} \\
\hline
 & Rel.\rule{5mm}{0mm}  & Nonrel.    & Rel.\rule{5mm}{0mm} & Nonrel.  \\
\hline
 $\surd\langle r^2 \rangle$ (GeV${}^{-1}$) & 3.177 & 3.263 & 3.235 & 3.320  \\
 $\langle r^{-1} \rangle$ (GeV) & 0.477 & 0.455 & 0.469 & 0.448  \\
 $\langle r^{-3} \ln mr \rangle$ (GeV${}^3$) & 0.495 & 0.448 & 0.489 & 0.443  \\
 $\langle r^{-3} \rangle$ (GeV${}^3$) & 0.504 & 0.414 & 0.496 & 0.406  \\
 $f_4(1P)$ (GeV${}^2$)& 0.1060 & 0.0783 & 0.1025 & 0.0748  \\
\hline
\end{tabular}
\end{center}

 $({}^a)$ For the parameters see Tab.~\ref{tab.01}
\end{table}

The relativistic corrections are becoming essential for some matrix
elements, which determine the fine structure splittings (see
Table~\ref{tab.02}). To calculate them in the relativistic case (for
the Salpeter equation) the expansion of the wave function in a series
over Coulomb-type functions was used as it was suggested in \cite{22}.
The numbers obtained have a computational error $\la 10^{-4}$  (the
dimension of the matrices D was varied from D=20 to D=40).

From the numbers given in Table~\ref{tab.02} one can conclude that

\begin{itemize}
\item for $1P$ and $2P$ states the root-mean-square radii practically coincide 
in the relativistic and the NR cases;
\item for the matrix element $\langle r^{-1}\rangle$  the difference between 
both cases is small, about 3\% for the $1P$ state and about 5\% for the $2P$ 
state; in the relativistic case $\langle r^{-1}\rangle$ and therefore 
$|a_{\rm NP}(nP)|$ is slightly larger.
\item in the relativistic case the values of $\langle r^{-3} \ln~mr \rangle$ are
about 7\% (10\%)  larger for the $1P(2P)$ state for given set of chosen
parameters;
\item for the Salpeter equation the matrix element $\langle r^{-3}\rangle$  is 
larger by about 14\% (22\%) for the $1P(2P)$ state;
\item the largest relativistic correction was found for the factor
$f_4$ given in Eq.~(\ref{eq.06}). This difference is about 30\% for the
$1P$ state and 36\% for the $2P$ state. The numbers given do practically not
change for different sets of parameters. So our averaged
value of $f_4(nP)$  $(\alpha_{\rm eff} \geq 0.35)$  are:  
\end{itemize}
\begin{equation}
f_4(1P) = 0.085 \pm 0.010~\;{\rm GeV}^2, \quad
f_4(2P) = 0.106 \pm 0.008  ~\;{\rm GeV}^2 .
 \label{eq.37}
\end{equation}
The theoretical error in Eq.~(\ref{eq.37}) ($\approx 10\%$)  mostly
comes from the variation of the $b$  quark mass (in the range $4.6 \div
5.0~\;{\rm GeV}$).

The increasing of $f_4(nP)$ in the relativistic case directly affects the values
of $\alpha_s(\mu)$  extracted from the fine structure data because according to
Eq.~(\ref{eq.17})
\begin{equation}
\alpha_s(\mu) = \sqrt{\frac{\pi m\Delta(nP)}{2f_4(nP)}},~~~~\Delta(nP)
=\eta_{\rm exp}(nP) - |a_{\rm NP}(nP)|,
 \label{eq.38}
\end{equation}
is proportional to $f_4^{-1/2}$ and $\alpha_s(\mu)$ is about 15\%
{\em smaller} in the relativistic case. This result obtains both for
$1P$ and $2P$ states.

Therefore, below we shall use only matrix elements calculated for the
Salpeter equation, in this way taking into account the relativistic
corrections. A last remark concerns the  choice of the quark pole mass,
$m_{\rm pole}=m$ which enters the Salpeter equation \cite{6}. Here we study the 
spin structure of the $\chi_b$  mesons determined by the spin-dependent
potentials now known only in one-loop approximation. Therefore the pole
mass of the $b$ quark will be taken also in one-loop approximation \cite{23}:
\begin{equation}
m=m_{\rm pole}=\bar{m}(\bar{m}^2) \left \{ 1 +
\frac{4}{3}\frac{\alpha_s(m_{\rm pole})}{\pi}\right \}
 \label{eq.39}
\end{equation}
In Eq.~(\ref{eq.39}) $\bar{m}(\bar{m}^2)$ is a running quark mass in the
$\overline{MS}$ renormalization scheme, its value from \cite{1} is
$\bar{m}=4.3 \pm 0.2$ GeV.  Then taking $\Lambda^{(4)}$  from
Eq.~(\ref{eq.19})  one finds $m$  in the range
\begin{equation}
4.5 ~\;{\rm GeV} \leq m \leq 5.0~\;{\rm GeV}
 \label{eq.40}
\end{equation}
Only values of the mass in this range will be used later in our calculations.

\section{$\alpha_s(\mu)$ for the $2P$ State \label{sec.8}}

For the $2P$ state $\alpha_s(\mu)$ can be immediately found from the relation (\ref{eq.17})
for the chosen static potential with fixed parameters $\alpha_{\rm eff}$,
$\sigma$, and $m$. At first, we shall give an estimate of $\alpha_s(\mu)$
using the following results:

\begin{enumerate}
\item The nonperturbative spin-orbit splitting $| a_{\rm NP}(2P)| $ depends
weakly on the choice of the parameters, provided the $b\bar{b}$  spectrum
is described with good accuracy
\begin{equation}
 | a_{\rm NP}(2P)  | =1.95 \pm 0.15 ~\;{\rm MeV} .
 \label{eq.41}
\end{equation}

\item In Eq.~(\ref{eq.15}) the experimental error of $\eta_{\rm exp}$
$(2P)$ is not large and therefore the difference $\Delta(2P)$
Eq.~(\ref{eq.17}) is known with an accuracy of about 20\%:
\begin{equation}
\Delta(2P) = \eta_{\rm exp}(2P) - | a_{\rm NP}(2P)| =  2.40 \pm 0.04
({\rm exp})\pm 0.15({\rm th})\; {\rm MeV}.
 \label{eq.42}
\end{equation}

\item In our calculations the matrix element $f_4(2P)$ is changing in the
narrow range:
\begin{equation}
f_4(2P) =  0.106 \pm 0.008~{\rm GeV}^2 .
 \label{eq.43}
\end{equation}
Then, from the fitting condition (\ref{eq.17})  and the numbers given
in Eqs.~(\ref{eq.41})-(\ref{eq.43}) the lower and upper bounds of
$\tilde{\alpha}_s(\mu)$ can be determined,
\begin{equation}
\sqrt{\frac{m}{m_0}} 0.37 \leq \tilde{\alpha}_s (\mu) \leq
\sqrt{\frac{m}{m_0}} 0.46
 \label{eq.44}
\end{equation}
Here a normalization mass, $m_0=\frac{1}{2}
M(\Upsilon(1S))=4.73~{\rm GeV}$ , was introduced for convenience. Here
and below all numbers were calculated in relativistic case, i.e. for
the Salpeter equation.
\end{enumerate}

From the estimates (\ref{eq.44})  it is clear that for the $2P$ state
$\alpha_s(\mu)\approx 0.40$  turns out to be large for any set of the
parameters of the static interaction. It is essentially larger than
that obtained in ref.~\cite{4} where $\alpha_s(3.25)=0.33$  and  in
ref.~\cite{5} where $\alpha_s(4.6)=0.26$. In our calculations large
values of $\tilde{\alpha}_s(\mu)$ are extracted
irrespectively to the value of the scale $\mu$, which is still not
fixed.

However, $\alpha_s(\mu)$  in Eq.~(\ref{eq.44})  is varying in a rather
wide range and its value is sensitive to small variations of the
factors entering the condition (\ref{eq.17}). The value of
$\alpha_s(\mu)$  is {\em decreasing} if the constant $\alpha_{\rm
eff}$  of the static interaction is growing. In our numerical
calculations the values of $\alpha_{\rm eff}$ is  supposed to be in the
range,
\begin{equation}
0.35 \leq \alpha_{\rm eff} < \alpha_B(q^2=0) \approx 0.48
 \label{eq.45}
\end{equation}
with a $b$ quark mass from the condition (\ref{eq.40}).

With the restriction (\ref{eq.45})  the fitted values of $\tilde{\alpha}_s(\mu)$
appeared to lie in the narrower range,
\begin{equation}
\tilde{\alpha}_s(\mu) = 0.40 \pm 0.02({\rm th})  \pm 0.04({\rm exp}) 
 \quad (b\bar{b}).
 \label{eq.46}
\end{equation}
Here the experimental error comes from $\eta_{\rm exp}$,
Eq.~(\ref{eq.18}), and the theoretical error is due to the variation of
$\alpha_{\rm eff}$, $m$, and $\sigma$.

In the extracted value $\tilde{\alpha}_s(\mu)$, Eq.~(\ref{eq.46}), the scale
$\mu$ is still not specified. To find $\mu_2$ it is better to
use the condition $c(2P)=c_{\rm exp}$,(\ref{eq.13}), for the
tensor splitting, because the theoretical uncertainty connected with the 
nonperturbative contribution to $c(2P)$  is negligible,
$c_{\rm NP} < 0.05$ Mev. This condition (\ref{eq.13}) turns out to be
satisfied for the scale,
\begin{equation}
 \mu =\mu_2 = 1.02 \pm 0.02 ~{\rm GeV},
 \label{eq.47}
\end{equation}
which has a small theoretical error, 2\%, while the
extracted value of $\tilde{\alpha}(\mu)$, Eq.~(\ref{eq.46}), was
determined with an accuracy of 15\%.

It is of interest to compare $\tilde{\alpha}(1.0)\approx 0.40$  with
the value found in perturbation theory. The scale $\mu_2\approx 1.0$
GeV is small, less than the running mass of the $c$ quark,
$\bar{m}_c=1.3\pm 0.2$ GeV \cite{1}, therefore $\alpha_s(1.0)$ should be
calculated with $\Lambda=\Lambda^{(3)}$  (2-{\rm loop}), $n_f=3$. The
value of $\Lambda^{(3)}$  can be found using the matching condition at
$\mu= \bar{m}_c$  and the value of $\Lambda^{(4)}$ (2-{\rm loop}), Eq.~(\ref{eq.19}).  Then
\begin{equation}
\Lambda^{(3)} (2-{\rm loop}) = 384^{+32}_{-30}  ~{\rm MeV}
 \label{eq.48}
\end{equation}
and correspondingly the two-loop strong coupling constant is
\begin{equation}
\alpha_s (1.0) = 0.53^{+0.6}_{-0.5},
 \label{eq.49}
\end{equation}
which is 30\%  larger than our fitted value given by
Eq.~(\ref{eq.46}).  It was suggested in \cite{10} that this decreasing
of $\tilde{\alpha}(\tilde{\mu})$  at the scale $\mu_2=1$ GeV can be
explained by the behavior of $\alpha_B(\mu)$ Eq.~(\ref{eq.27}) in the
vacuum background field, thus demonstrating the phenomenon of freezing
of $\alpha_s(\mu)$ . In \cite{10}, from a fit to the charmonium fine
structure, the background mass $m_B$ in Eq.~(\ref{eq.27})  was found to
be (in the $\overline{MS}$ renormalization scheme) 
\begin{equation}
m_B=1.1~{\rm GeV}, \quad \Lambda^{(3)}_B(2-{\rm loop})=400^{+40}_{-50} \;
 {\rm MeV} \quad (c\bar{c}).
 \label{eq.50}
\end{equation} 
Our extracted value of $\tilde{\alpha}(1.0)$  in
Eq.~(\ref{eq.46})  corresponds to the close value of $\Lambda^{(3)}_B$,
\begin{equation}
\Lambda^{(3)}_B (2-{\rm loop}) = 420^{+40}_{-30}\; {\rm MeV} \quad(b\bar{b}).
 \label{eq.51}
\end{equation}
Note also that for the $1P$ state in charmonium the value
\begin{equation}
\tilde{\alpha}(1.0) = 0.38 \pm 0.03 ({\rm th})  \pm 0.04  ({\rm exp})
 \label{eq.52}
\end{equation}
practically coincides with $\tilde{\alpha}(1.0)$  in bottomonium,

\begin{equation}
\tilde{\alpha}(1.0) = 0.40 \pm 0.02 ({\rm th})  \pm 0.04
({\rm exp}),\quad \mu_2 = 1.02\pm 0.02~{\rm GeV} .
 \label{eq.53}
\end{equation}
This coincidence is not, in our opinion, accidental: both states, the $c\bar{c}
$ $1P$ state and the $b\bar{b}$ $2P$  state, have the same size:
$R=\sqrt{<r^2>_{nP}}=0.62 \div 0.65$ fm. This  coincidence of the values of 
$\alpha_s(\mu)$ and of the sizes indicates that for the bound states the
scale $\mu$ is characterized by the size, but not the momentum, of the system.
This result is in agreement with the predictions of refs. \cite{6,7}.

With the use of the fitted values $\tilde{\alpha}_s(\mu_2)$,
Eq.~(\ref{eq.53}), the theoretical number obtained for
the spin-orbit splitting $a_{\rm tot}$ automatically satisfies the
third fitting condition Eq.~(\ref{eq.13}). Calculated numbers of $a$
and $c$  are given in Table~\ref{tab.04} for three different sets of 
parameters. From these numbers  one can see that the second order
radiative corrections $a^{(2)}_{\rm P}$ and $c^{(2)}_{\rm P}$ are negative and
rather large: about 25\%  for the tensor and 40\%  for the spin-orbit
splittings.

\begin{table}
\caption[]{Fine-structure parameters for the $2P$ $b\bar{b}$ state.}
\label{tab.04}
\vspace{1ex}
\begin{center}
\begin{tabular}{|l|r|r|r|r|}
\hline
 & Set I $({}^a)$  & Set II $({}^a)$ & Set III $({}^b)$ & Exp. val. \\
\hline
 $\tilde{\alpha}(\mu_2)$  & 0.392 & 0.429 & 0.386 &   \\
 $\mu_2$ (GeV) & 1.03  & 1.02  & 1.03  &   \\
 $c^{(2)}_{\rm P}$ (MeV) & $-2.62$ & $-3.14$ & $-3.35$ &   \\
 $c_{\rm tot}$ (MeV) & 9.12  & 9.11  & 9.17  & $9.1\pm0.2$ \\
 $a_{\rm NP}$ (MeV) & $-2.12$ & $-1.83$ & $-1.80$ &   \\
 $a^{(1)}_{\rm P}$ (MeV) & 17.61 & 18.33 & 18.77 &   \\
 $a^{(2)}_{\rm P}$ (MeV) & $-6.12$ & $-7.19$ & $-7.52$ &  \\
 $a_{\rm tot}$ (MeV) & 9.37  & 9.32  & 9.45  & $9.4\pm0.2$ \\
\hline
\end{tabular}
\end{center}

 $({}^a)$ For the parameters see Tab.~\ref{tab.01}\\
 $({}^b)$ $\alpha_{\rm eff} = 0.386$, $\sigma = 0.185$ GeV${}^2$, $m=5.0$ GeV.
\end{table}

Note that we have met here no difficulty to get a precise description
of the tensor and spin-orbit splittings for the $2P$ state simultaneously, 
in contrast to the results of
ref.  \cite{5}, where some difficulties have occurred, in our opinion, because
of the choice $\tilde{\mu}=m$ (see the discussion in Section \ref{sec.5}).

\section{ $\alpha_s(\mu)$  for the $1P$ State \label{sec.9}}

For the $1P$ state the scale-independent condition (\ref{eq.17})
cannot be used directly, because the important factor $\Delta(1P)$ in
Eq.~(\ref{eq.17}) has a large experimental error. So in this case one
needs to use the two $\mu$-dependent conditions, Eq.~(\ref{eq.13}), on
the splittings $a$ and $c$.

There exist a lot of variants where these two conditions can be
satisfied.  However, in many cases the two-loop values
$\tilde{\alpha}_s(\mu_1)$ and $\mu_1$, extracted from those fits,
correspond to a very large value of the QCD constant $\Lambda^{(4)}$.
Therefore, the additional requirement (21) that $\Lambda^{(4)}$ 
(2-{\rm loop}) should have a value  in the range 
307 MeV $\leq \Lambda{(4)}\leq 371 $MeV, is necessary. 
If this restriction is put, then in our
calculations the extracted scale $\mu_1$  appears to lie in the
narrow range,
\begin{equation}
 \mu_1 = 1.80 \pm 0.10 ~{\rm GeV}
 \label{eq.54}
\end{equation}
and
\begin{equation}
 \tilde{\alpha}(\mu_1) = 0.33 \pm 0.01({\rm exp}) \pm 0.02 ({\rm th}) .
 \label{eq.55}
\end{equation}

Our value for the scale $\mu_1$  turned out to be very close to that
determined in refs. \cite{7}, but our fitted value of $\alpha_s(\mu_1)$
is about 15\% larger than the one found in \cite{7} where
$\tilde{\alpha}_s$(3-{\rm loop})=0.29 and $\Lambda^{(4)}(3-{\rm
loop})=230$ MeV (or $\Lambda^{(4)}$ (2-{\rm loop})$=250^{+90}_{-60}$
MeV) is smaller than in our fit.

For the $1P$ state it was also observed that if a large value
$\sigma=0.2$ GeV$^2$  is taken, then it is difficult to reach a
consistent description of the tensor and the spin-orbit splittings
simultaneously. Therefore here, as well as in the charmonium case
\cite{10}, the values $\sigma=0.17\div 0.185$ GeV$^2$  are considered
as preferable. Also the choice of a relatively large $b$  quark mass,
\begin{equation}
 m_b = 4.75 \div 4.9 ~{\rm GeV},
 \label{eq.56}
\end{equation}
gives rise to a better fit.

\begin{table}
\caption[]{Fine-structure parameters for the $1P$ $b\bar{b}$ state.}
\label{tab.05}
\vspace{1ex}
\begin{center}
\begin{tabular}{|l|r|r|r|r|}
\hline
 & Set I $({}^a)$  & Set II $({}^a)$ & Set III $({}^b)$ & Exp. val. \\
\hline
 $\tilde{\alpha}(\mu_2)$  & 0.335 & 0.340 & 0.32  &   \\
 $\mu_1$ (GeV) & 1.80  & 1.85  & 1.90  &   \\
 $c^{(2)}_{\rm P}$ (MeV) & $0.96$ & $1.03$ & $0.90$ &   \\
 $c_{\rm tot}$ (MeV) & 11.93  & 11.92  & 11.91  & $11.92\pm0.20$ \\
 $a_{\rm NP}$ (MeV) & $-2.82$ & $-2.45$ & $-2.44$ &   \\
 $a^{(1)}_{\rm P}$ (MeV) & 16.46 & 16.34 & 16.52 &   \\
 $a^{(2)}_{\rm P}$ (MeV) & $0.12$ & $0.21$ & $0.05$ &  \\
 $a_{\rm tot}$ (MeV) & 13.76 & 14.09 & 14.12 & $14.23\pm0.57$ \\
\hline
\end{tabular}
\end{center}

 $({}^a)$ For the parameters see Tab.~\ref{tab.01}\\
 $({}^b)$ $\alpha_{\rm eff} = 0.386$, $\sigma = 0.185$ GeV${}^2$, $m=5.0$ GeV.
\end{table}

The results of our calculations for the $1P$ state are given in
Table~\ref{tab.05} from which one can see that the second order corrections
$c^{(2)}_{\rm P}$ and $a^{(2)}_{\rm P}$ are relatively small, 8\% and 15\%, but
still remain important for a fit to the experimental data. Also in all
good fits the effective Coulumb constant $\alpha_{\rm eff}$  lies
between $\tilde{\alpha}(\mu_1)$  and $\tilde{\alpha}_s(\mu_2)$:
\begin{equation}
\tilde{\alpha}(\mu_1) < \alpha_{\rm eff} \leq \tilde{\alpha}(\mu_2) .
 \label{eq.57}
\end{equation}

In our analysis $\mu_2(2P)$ is less than $\mu_1(1P)$  and their ratio
is almost inversely proportional to the ratio of the radii of these states:
\begin{equation}
\frac{\mu_1(1P)}{\mu_2(2P)} \approx 1.7 \div 1.8; \quad
\frac{\sqrt{\langle r^2 \rangle_{2P}}}{\sqrt{\langle r^2 \rangle_{1P}}}= 1.6
\div 1.65.
 \label{eq.58}
\end{equation}
This result is in full agreement with the prediction of refs.~\cite{6,7} about
the decrease of the scale with increasing principal quantum number.

\section{Conclusion \label{sec.10}}

The precise experimental data on the masses of $\chi_{bJ}(1P)$ and
$\chi_{bJ}(2P)$ give a unique opportunity to determine the QCD strong
coupling constant at low-energy scales. In our analysis of fine
structure splittings we found that:

\begin{enumerate}
\item The relativistic corrections which are small for such characteristics
as the $b\bar{b}$ levels, radii, and matrix element $\langle r^{-1} \rangle$,
are nevertheless essential for the determination of the factor
$f_4(nP)$, which is inversely proportional to the extracted value of
$\tilde{\alpha}^2_s(\mu)$.

\item  From a $\mu$-independent analysis of the $2P$ state,
the value  $\tilde{\alpha}_s(\mu_2)\approx 0.40$  was extracted.
The scale $\mu_2=1.0\pm 0.02$ GeV, determined from the tensor
splitting, appeares to be practically unchanged for any chosen set of
parameters.

\item The extracted value $\tilde{\alpha}(1.0)\approx 0.40$  is about 30\%
lower than the one found in perturbation theory if
$\Lambda^{(3)}=384^{+32}_{-30}$ MeV was used.  This value agrees with
the fitted $\alpha_s(1.0, c\bar{c})$ extracted from the analysis of the
charmonium fine structure. This result  can be naturally explained in
the framework of background perturbation theory.

\item The scale $\mu_1\approx 1.8$ GeV for the $1P$ $b\bar{b}$  state obtained here
agrees with the prediction in \cite{7} but corresponds to the larger value
$\Lambda^{(4)}$(2-{\rm loop})=$338^{+33}_{-31}$ MeV, which gives rise to
$\alpha_s(M_z)=0.119\pm 0.002$.

\item The preferred values of the pole mass of the $b$ quark are found to be
$m=4.7\div 4.9$ GeV but from the fine structure analysis we could not 
narrow their range.
\end{enumerate}

Our results have confirmed the important observation of Yndurain et al.
\cite{6,7} that the strong coupling constant is increasing for states
with a larger size or larger principal quantum number and this fact is
essential in many aspects of quarkonium physics.

\vspace{10mm}

One of the authors (A.M.B.)  acknowledges a partial financial support
through the grants RFFI-DFG No.96-02-00088g


\newpage
\begin{thebibliography}{99}
\bibitem{1} Data compilations: Particle Data Group, The European
Physical Journal {\bf C 3}, 1, (1998)
\bibitem{2} W. Buchm\"uller, Y.J. Ng and S.-H.H. Tye, Phys. Rev. {\bf D24},
3003 (1981);
W. Buchm\"uller and S.-H.H. Tye, Phys. Rev. {\bf D24}, 132 (1981)
\bibitem{3} J. Pantaleone, S.-H.H. Tye, and Y.J. Ng, Phys. Rev. {\bf D33},
777 (1986)
\bibitem{4} L.P. Fulcher, Phys. Rev. {\bf D44}, 2079 (1991)
\bibitem{5} F. Halzen, C. Olson, M.G. Olsson, and M.L. Stong, Phys. Rev.
{\bf D47}, 3013 (1993)
\bibitem{6} S. Titard and F.J. Yndurain, Phys. Rev. {\bf D49}, 6007 (1994);
{\bf 51} 6348 (1995)
\bibitem{7} S. Titard and F.J. Yndurain, Phys. Lett. {\bf 351}, 541 (1995);
A. Pineda, Phys. Rev. {\bf D55}, 407 (1997)
A. Pineda and F.J. Yndurain, Phys. Rev. {\bf D58}, 094022 (1998)
\bibitem{8} S. Narison, Nucl. Phys. Proc. Suppl. {\bf 54A}, 238 (1997)
\bibitem{9} K. Igi and S. Ono, Phys. Rev. {\bf D36}, 1550 (1987);
{\bf D37}, 1338E (1988)
\bibitem{10} A.M. Badalian and V.L. Morgunov, Phys. Rev. {\bf D60} 116008
 (1998); hep-ph/9901430
\bibitem{11} Yu.A. Simonov, Phys. Usp. {\bf 39}, 313 (1996); hep-ph/9709344
\bibitem{12} Yu.A. Simonov, Phys. Rep. {\bf 320}, 265 (1999)
\bibitem{13} F.J. Yndurain, hep-ph/9910399; 
 Nucl. Phys.(Proc.Suppl) {\bf B64}, 433 (1998)
\bibitem{14} Yu.A. Simonov, Pis'ma JETP, {\bf 57}, 513 (1994); 
Yad. Fiz. {\bf 58}, 113 (1995)
\bibitem{15} A.M. Badalian, Phys. At. Nucl. {\bf 60}, 1003 (1997);
M.N. Chernodub, F.V. Gubarev, M.I.Policarpov, and V.I.Zakharov,
hep-ph/0003006 (submitted to Phys.Lett)
\bibitem{16} A.M. Badalian and Yu.A. Simonov, Phys. At. Nucl. {\bf 59}, 2164
(1996);
Yad. Fiz. {\bf 59}, 2247 (1996)
\bibitem{17} A. Di Giacomo and H. Panagopoulos, Phys. Lett. {\bf B285},
133 (1992);
M. D'Elia, A. Di Giacomo, and E. Maggiolaro, Phys. Lett. {\bf B408}, 315 (1997)
\bibitem{18} A.M. Badalian and V.P. Yurov, Yad. Fiz. {\bf 56}, 239 (1993);
 Phys. Atom. Nucl. {\bf 56}, 1760 (1993)
\bibitem{19} A.M. Badalian and V.L. Morgunov, Phys. At. Nucl. {\bf 62},
1019 (1999); 
Yad. Fiz. {\bf 62}, 1086 (1999)
\bibitem{20} M. Peter, Phys. Rev. Lett. {\bf 78}, 602 (1997); 
 Y. Schr{\"o}der, hep-ph/9812205 (1998); 
 Phys. Lett. {\bf B447}, 321 (1999)
\bibitem{21}  A.M. Badalian and Yu.A. Simonov, Phys. At. Nucl. {\bf 60}, 630
(1997); 
 Yad. Fiz {\bf 60}, 714 (1997)
\bibitem{22}  S. Jacobs, M.G. Olsson, and C. Suchyta, Phys. Rev. {\bf D33}, 3338
(1986)
\bibitem{23} F.J. Yndurain, {\em The Theory of Quark and Gluon Interactions},
3rd edition, (Springer, 1999)

\end{thebibliography}
\end{document}